# An Internet Multicast System for the Stock Market

*N. F. Maxemchuk*
*D. H. Shur*

AT&T Labs - Research

*ABSTRACT*

We are moving toward a distributed, international, twenty-four hour, electronic stock exchange. The exchange will use the global Internet, or internet technology. This system is a natural application of multicast because there are a large number of receivers that should receive the same information simultaneously.

The data requirements for the stock exchange are discussed. The current multicast protocols lack the reliability, fairness, and scalability needed in this application. We describe a distributed architecture together with a reliable multicast protocol, a modification of the RMP protocol, that has characteristics appropriate for this application.

The architecture is used in three applications:

— In the first, we construct a unified stock ticker of the transactions that are being conducted on the various physical and electronic exchanges. Our objective is to deliver the the same combined ticker reliably and simultaneously to all receivers, anywhere in the world.

— In the second, we construct a unified sequence of buy and sell offers that are delivered to a single exchange or a collection of exchanges. Our objective is to give all traders the same fair access to an exchange independent of their relative distances to the exchange or the loss characteristics of the international network.

— In the third, we construct a distributed, electronic trading floor that can replace the current exchanges. This application uses the innovations from the first two applications to combine their fairness attributes.

## 1. Introduction

An exchange is any organization, association or group which provides or maintains a marketplace where securities, options, futures or commodities can be traded. There are hundreds of exchanges around the world. Yahoo lists 107 stock exchanges.

Traditional stock exchanges are centralized. The exchange is located in a single physical place and all data (both market and trades) flows through a single system. The centralized system is responsible for transaction reporting, and settlements for the purpose of exchanging assets.

Electronic exchanges, like the NASDAQ, allow remotely located traders to connect with the centralized system. The traders send and receive information, trades, etc., over dedicated access lines and private networks. Connections are expensive, and thus limit who can be connected.

The Internet now allows, or is on the verge of allowing, almost anyone to connect to almost any electronic exchange, in any country, at very low cost. Internet communication technology is enabling the restructuring of the stock exchanges and a number of Internet based private stock trading systems are emerging. Private stock trading systems interface into other trading systems, such as the NASDAQ. The private system may satisfy buy and sell requests among their own subscriber base, or may pass on the trade to the larger exchange.

The electronic stock market presents opportunities, as noted in these excerpts from the prepared remarks by Frank G. Zarb, Chairman and CEO of the National Association of Securities Dealers Inc. before the National Press Club in Washington, D.C. on Wednesday, June 23, 1999:

> *In a very few years, trading securities will be digital, global, and accessible 24 hours a day.*

> *People will be able to get stock price quotations instantly and instantly execute a trade any time of day or night, anywhere on the globe, with stock markets linked and almost all-electronic.*

> *As for stock markets, they will see global alliances, mergers, and new electronic ventures. That will give companies listed on these markets access to pools of capital internationally, not just domestically, and consumers will be able to invest in a worldwide list of companies as easily as trading locally.*

> *This 21st century stock market will be multi-dealer, computer-screen based, technology-driven and open to all - all because people will have access to information that they want to act on.*

An electronic exchange also creates technical challenges, as noted in these excerpts from a New York Times article on Sept. 23, 1999, of an interview with Arthur Levitt, the chairman of the Securities and Exchange Commission:

> *Levitt seems most concerned that if trading continues to migrate to the new electronic market systems, investors may not get the best prices. Information about orders and transactions across the entire market are not now available in one place. Technology, he said, allows the creation of a central system in which investors will be fully informed about prices everywhere, from the New York Stock Exchange, the NASDAQ, the American Stock Exchange and the new systems.*

> *He stressed that he was asking for the development of a technology that would allow all orders to be shown to investors, not an institution or place where all orders would be executed.*

We are proposing a distributed architecture and protocol which can meet these challenges, and more. The architecture is used to implement three applications. In this document we focus on stock exchanges although many of the ideas are applicable to other types of exchanges and the fairness attributes are appropriate for some distributed games.

In the first application, the trades that are executed on hundreds of trading floors, distributed globally, are merged into a single information stream and presented to millions or tens of millions of investors, internationally. Our objective is to provide the same information stream, with all of the reports in the same order, to all of the investors at the same time, no matter where in the world they are, or where the trading floors are located. We meet this objective using a modified version of the reliable multicast protocol, RMP[1]. We show that our architecture does not have the scaling problems that are common in many reliable multicast applications.

In the second application, the buy and sell orders from millions of investors are merged into a single stream and presented to a single exchange, or simultaneously presented to several exchanges. The exchanges operate under their own rules. Currently, investors that are in the same city as an exchange have an advantage over investors on the other side of the world. If both investors see a ticker at the same time and submit a bid, the bid from the closer investor may reach the exchange several seconds earlier on the average. Our objective is to give all investors fair access to the exchange. This application differs from the first mostly in the cryptographic techniques that must be used. The cryptographic techniques, and the trust relationships, will be explained in section 6, after the architecture is described in section 4.

In the third application we use a distributed rule to operate on the buy, sell, and cancel orders to create a distributed exchange. This exchange makes use of the sequencing property of RMP, and also the characteristic that every receiver eventually knows that every other receiver has seen the same sequence. This distributed exchange can operate under several rules for executing trades. Distributed exchanges can be among the trading floors in the first two applications.

## 2. RMP

The RMP protocol was originally implemented in 1983 to build a distributed database, with redundant data, on an Ethernet[2]. The computers on the Ethernet independently lost messages due to buffer overflows caused by competing processes. The protocol predated the word "multicast" and the multicast implementation on the Internet[3]. It was originally called the reliable broadcast protocol, RBP. RBP has three characteristics that distinguishes it from earlier protocols.

The first characteristic of RBP is that it guarantees that every receiver places the messages in the same sequence. Redundant databases are simpler to implement when the programs running on the different machines always place the database in the same state. RBP's message sequencing preserves the database state.

The second characteristic of RBP is that every receiver eventually knows that every other receiver has the data. The primary reason for redundant data is to be able to operate during failures that make some copies of the data unavailable. By only committing database operations that can be performed on every active copy of the database, we guarantee data consistency when copies of the database fail and are restored.

The third characteristic of RBP is that it uses only one control message per data message, independent of the number of receivers, when there aren't any losses. The number of messages that are transmitted when there are losses is derived in reference 4. The number of control messages should be compared with earlier reliable protocols, that require at least as many control messages as receivers when there aren't any losses, without providing the first two characteristics.

A modified version of RBP was applied to the Internet multicast network and called RMP[5]. NASA maintains a WEB site for recent work on RMP at http://research.ivv.nasa.gov/RMP/.

### 2.1 The Original Protocol

Before describing how RMP is modified to meet the requirements of the stock exchange, it is necessary to understand its original operation. The protocol has two parts. The first part operates on multicast messages during normal operation. It guarantees delivery and ordering of the messages from the sources. The second part is a reformation protocol that reorganizes the broadcast group and guarantees database consistency after failures and recoverys. In this paper we are only concerned with the first part of the protocol. The original reformation protocol is based on a three phase commit procedure and is completely distributed. We use a simpler, centralized procedure in this application.

There are $n$ sources and $m$ receivers, as shown in figure 1. The sources and receivers may be the same or different. The objective is for the $m$ receivers to place the messages from the $n$ sources in the same order, regardless of which receivers fail to receive which messages. This is accomplished by having a single receiver acknowledge messages. The acknowledgement assign the message a sequence number and all of the receivers place the messages in the order indicated by the sequence number. We guarantee that every receiver has all of the messages by sequentially passing the responsibility to acknowledge messages to each receiver and requiring a receiver to acquire all of the preceding messages before acknowledging a new message.

A message from source $s$ contains the label $(s, M_s)$ to signify that it is the $M_s^{th}$ message from source $s$. Source $s$ transmits message $M_s$ at regular intervals until it receives an acknowledgement or decides that the token site is not operating. When a source decides that the token site is not operating it initiates the reformation process to form a new token group.

The receivers take turns acknowledging messages from sources by passing a token. Each of the receivers is assigned a unique number from 0 to $m-1$. When the token site at receiver number $r$ sends an acknowledgement, the message serves three separate functions:

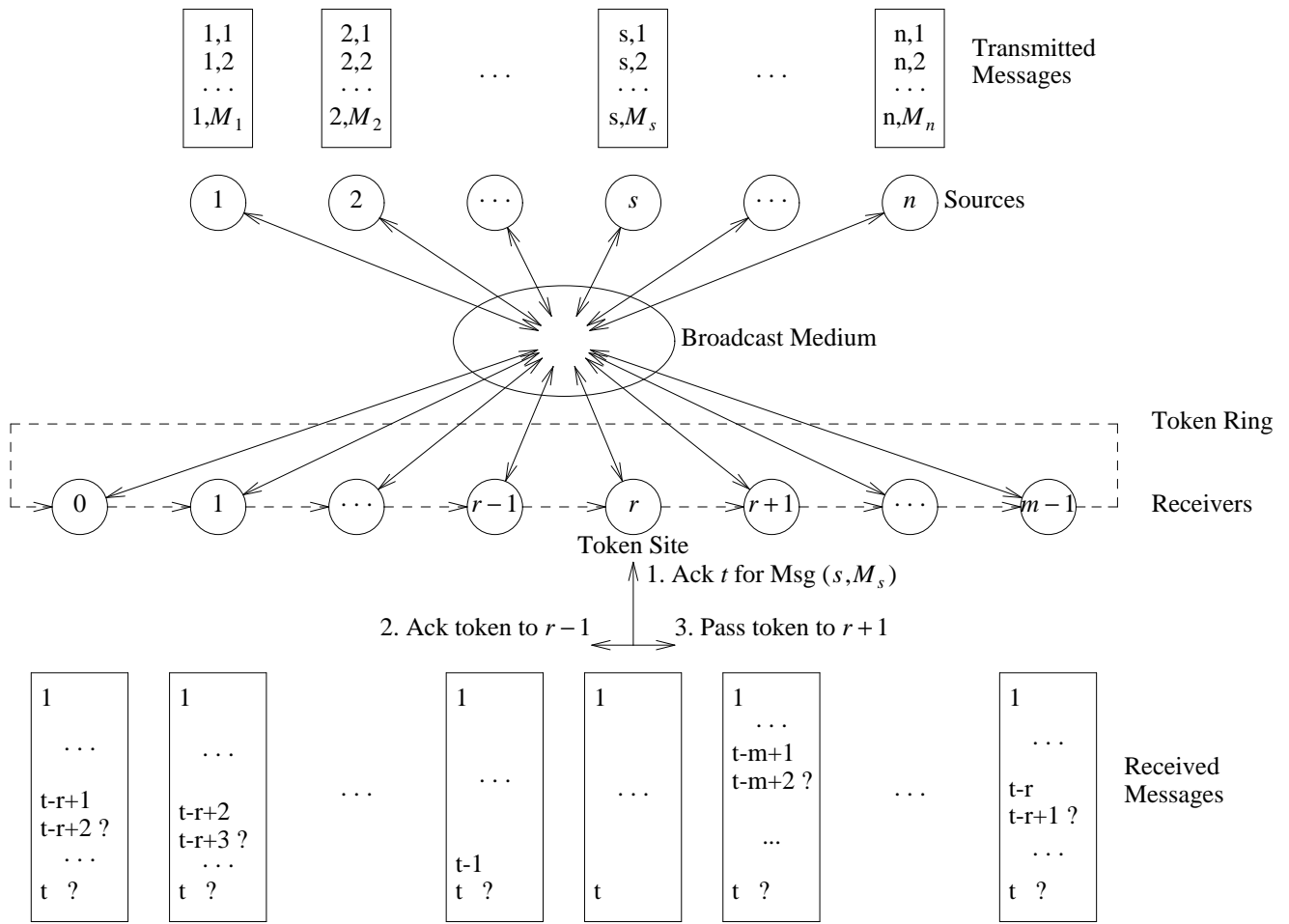

**Figure 1.** The Reliable Broadcast Protocol

1. The message acknowledges $(s, M_s)$ and assigns it global sequence number $t$.

2. The message is an acknowledgement to the previous token site, receiver $\lceil r-1 \rceil$ mod $m$, that the token was successfully transferred to $r$.

3. The message transfers the token to the next token site, receiver $\lceil r+1 \rceil$ mod $m$.

Token site $r$ periodically sends acknowledgement $t$ until it receives acknowledgement $t+1$, which acknowledges that receiver $\lceil r+1 \rceil$ mod $m$ accepted the token. If an acknowledgement isn't received in a specified number of attempts, receiver $r$ decides that receiver $r+1$ is inoperable and initiates a reformation process. In order to prevent unnecessary reformations, receiver $r+1$ transmits a special token acknowledgement when there are no source messages to acknowledge.

As soon as $r$ sends acknowledgement $t$ it gives up the right to acknowledge new source messages, even though it is not certain that $\lceil r+1 \rceil$ mod $m$ has received the token. This guarantees that at most one receiver acknowledges source messages.

Receiver $\lceil r+1 \rceil$ mod $m$ does not accept the token transferred by acknowledgement $t$ until it has all of the acknowledgements and source messages that were acknowledged up to and including $t$. When a receiver accepts the token it also assumes responsibility for servicing retransmission requests for

— acknowledged source messages that other receivers failed to receive,

— acknowledgements for the source messages that either the source or one of the other receivers have failed to receive, and

— acknowledgements for token passing messages that a previous token site failed to receive.

Retransmission requests from other receivers are explicit. A receiver requests missing messages when it receives an acknowledgement for a source message $(s,M_s)$ that has not been received, an acknowledgement with a larger sequence number $t$ than expected, or a message from source $s$ with a larger $M_s$ than expected. Retransmission requests from sources and previous token sites are implicit. If a source retransmits a message that has been acknowledged, the retransmission server assumes that the source failed to receive the acknowledgement. If a previous token site retransmits a token passing message, the retransmit server assumes that it failed to receive the token passing acknowledgement. Receiver $r$ (the last token holder) does not stop servicing retransmission requests until it receives the acknowledgement for passing the token. This guarantees that there is at least one site that is responding to retransmission requests. In particular, if receiver $r+1$ does not have all of the requisite messages to accept the token, it can recover them from $r$ before accepting the token.

Receivers process arriving messages and acknowledgements and place the messages in the sequence indicated by the sequence number of the acknowledgements. Each receiver, $r$, tracks $t_r$ and $M_{s,r}$ for each $s$, where $t_r$ is the next acknowledgement number that $r$ expects, and $M_{s,r}$ is the next message that $r$ expects to process from source $s$. Receiver $r$ has all of the acknowledgements below $t_r$. Receiver $r$ has received acknowledgements for all $M_s < M_{s,r}$.

The receivers use a negative acknowledgement strategy: No control messages are sent unless a missing acknowledgement or source message is detected. When a receiver detects a missing message, it periodically requests the message until it receives the message or decides that the retransmit server is inoperable and initiates a reformation process.

When $r$ receives source message $(s,M_s)$:

if: $M_s < M_{s,r}$, $(s,M_s)$ is a duplicate.

if: $M_s = M_{s,r}$,

if: $(s,M_s)$ is not in the store of waiting messages, it is placed in the store waiting for acknowledgement.

otherwise: $(s,M_s)$ is a duplicate.

if: $M_s > M_{s,r}$, The acknowledgement for $(s,M_s-1)$ is missing, and is requested.

When $r$ receives acknowledgement $t$ for $(s,M_s)$,

if: $t < t_r$, acknowledgement $t$ is a duplicate.

if: $t = t_r$,

if: $(s,M_s)$ has been received, the message is placed in position $t$ in the receive list and $t_r$ and $M_{s,r}$ are incremented.

otherwise: $(s,M_s)$ is missing and is requested before processing acknowledgement $t$.

if: $t > t_r$, acknowledgement $t-1$ is missing and is requested before processing acknowledgement $t$.

As the token is passed, RMP guarantees that every receiver eventually receives the acknowledged messages and that every receiver eventually knows that every other receiver has received these messages. When a new acknowledgement is transmitted, the receiver that acknowledged that message has that message and all of the acknowledged source messages that preceded it. We can also infer that the previous token sites had all of the messages that they required to send their acknowledgements. For instance, when acknowledgement $t$ is transmitted from receiver $r$:

— receiver $r$ has all of the acknowledged messages up to and including the $t^{th}$ message,

— receiver $\lceil r-1 \rceil \bmod m$ has all of the acknowledged messages up to and including the $(t-1)^{th}$ message,

— $\cdots$, and that

— receiver $\lceil r-m+1 \rceil \bmod m$ has all of the acknowledged messages up to and including the message acknowledged by $t-m+1$.

Since $\lceil r-m \rceil \bmod m = r$, when the $t^{th}$ acknowledgement is transmitted all of the other receivers have all of the source messages up to and including the message acknowledged by $t-m+1$.

Even if none of the other receivers receive the $t^{th}$ acknowledgement, when the $t^{th}$ acknowledgement is transmitted receiver $r$ knows that all of the receivers have all or the messages up to and including $(t-m+1)^{th}$ message. In addition,

— receiver $\lceil r-1 \rceil \bmod m$ knows that all receivers have all of the messages up to and including the $(t-m)^{th}$ message,

— receiver $\lceil r-2 \rceil \bmod m$ knows that all receivers have all of the messages up to and including the $(t-m-1)^{th}$ message,

— $\cdots$, and that,

— receiver $\lceil r-m+1 \rceil$, and therefore, all of the receivers know that all of the receivers have all of the messages up to and including the $(t-m+2)^{th}$ message.

When there are no new source messages to acknowledge in a negative acknowledgement strategy, receivers that missed the last source messages or acknowledgements do not detect their loss. We can provide reception guarantees in bounded time by periodically passing the token when there are no new source messages to acknowledge.

### 3. Requirements of the Application

The requirements of a global stock market are very different from those of a replicated database on a local area network. Some of the issues that are caused by these differences are addressed by the original RMP protocol. However, other issues require modifications of the RMP protocol and architecture. We will state the differences, then describe the modifications that are needed to address them.

In the stock market applications:

1. There are a large number of receivers that join and leave the applications frequently. This is particularly true in the first application, where we distribute a unified ticker to millions of stock traders.

   — A network with millions of receivers must pass the token millions of times before the guarantees that are available with RMP are realized.

   — If receivers enter or join the group frequently, the protocol will spend most of its time reorganizing the receiver list, rather than ordering messages.

   — Most messages will be lost by some receivers and have to be recovered.

2. The sources and receivers in the stock market cannot be trusted to the extent that they are trusted in a distributed database.

   — It is unreasonable to expect individuals who are competing in a stock market to cooperate to recover missed messages and or to provide fair access to one another.

3. In many of the stock market applications data quickly becomes obsolete. A new trading price replaces the price from a few seconds earlier.

4. The number of messages in a unified stock ticker is much greater than in a local, replicated database.

   — There may be periods when the average arrival rate of messages will be greater than the average time that it takes to pass a token. During these periods the queue of messages waiting to be acknowledged increases. If the high arrival rate persists, the backlog of unacknowledged messages increases and may result in unacceptable delays.

   — The amount of data in the composite ticker may exceed the amount of data that can pass through the receivers in the repair group.

   — The amount of data will frequently exceed the bit rate that can be transmitted to some of the end users. The system should not be constrained by the least capable receiver.

5. The data should be delivered to all of the recipients simultaneously so that they all have the same opportunity to act on the information.

   — The delivery time should not depend on the location of the receiver. A trader that lives in the same city as a trading floor should receive the ticker at the same time as a trader in a different part of the world - even though the difference in propagation delay alone may exceed 100 milliseconds.

   — The delivery time should be independent of the probability of message loss between different locations and the time that it takes to recover the missing messages.

6. Every source should have an equal opportunity to have its data received promptly. This should hold even though some sources are near the receiver, and have few message losses and others are far and must retransmit several times.

   — RMP achieves transmitter fairness by rotating the site that acknowledges messages among a group of receivers that are distributed around the world. In effect, the entry portal for the sequence of acknowledged messages moves so that the transmitters that have the best access changes.

7. Security concerns must be addressed, including:

   — constraining transmission access to authorized sources,

   — preventing early reception of the data stream,

   — limiting reception to authorized receivers,

   — preventing spoofing or adding to the repaired sequence, and

   — preventing denial of service attacks.

8. The system must quickly detect and adjust to the failure of any components.

   — The token passing mechanism guarantees that repair group failures are detected within the time it takes the token to cycle around the repair group. However, in the original protocol the cycle time is dependent on the number of receivers and the average arrival rate of messages.

9. The messages that are lost by the receivers on a multicast network on the Internet are correlated.

   — The multicast network is physically a tree network. When a message is lost or corrupted at a node of the tree, all of the receivers that are fed from that node lose the same message.

   — When many receivers lose the same message there may be a NACK implosion[6,7,8,9].

## 4. Architecture

The scaling and trust issues in the stock market application lead to a change in the basic architecture. The architecture is multi-layered, and the inner layer is constrained to trusted components. The quantity of data transmitted in this application results in a replication of parts of the architecture.

All three of the stock market applications use the multicast architecture that is shown in figure 2. The reliable multicast protocol, RMP, operates in the core of the architecture on a global multicast tree. The layered architecture maps the large number of sources or receivers in an application onto a more manageable numbers that operate in the core of the protocol. Specifically secondary sources are aggregated at primary sources, and primary and secondary receivers distribute data customers as depicted in figure 2. RMP can operate with less than a few hundred receivers and less than a few thousand sources, even though the stock market applications may have millions of sources and receivers. The scaling issues associated most reliable multicast protocols are avoided by the architecture, rather than being resolved by the protocol.

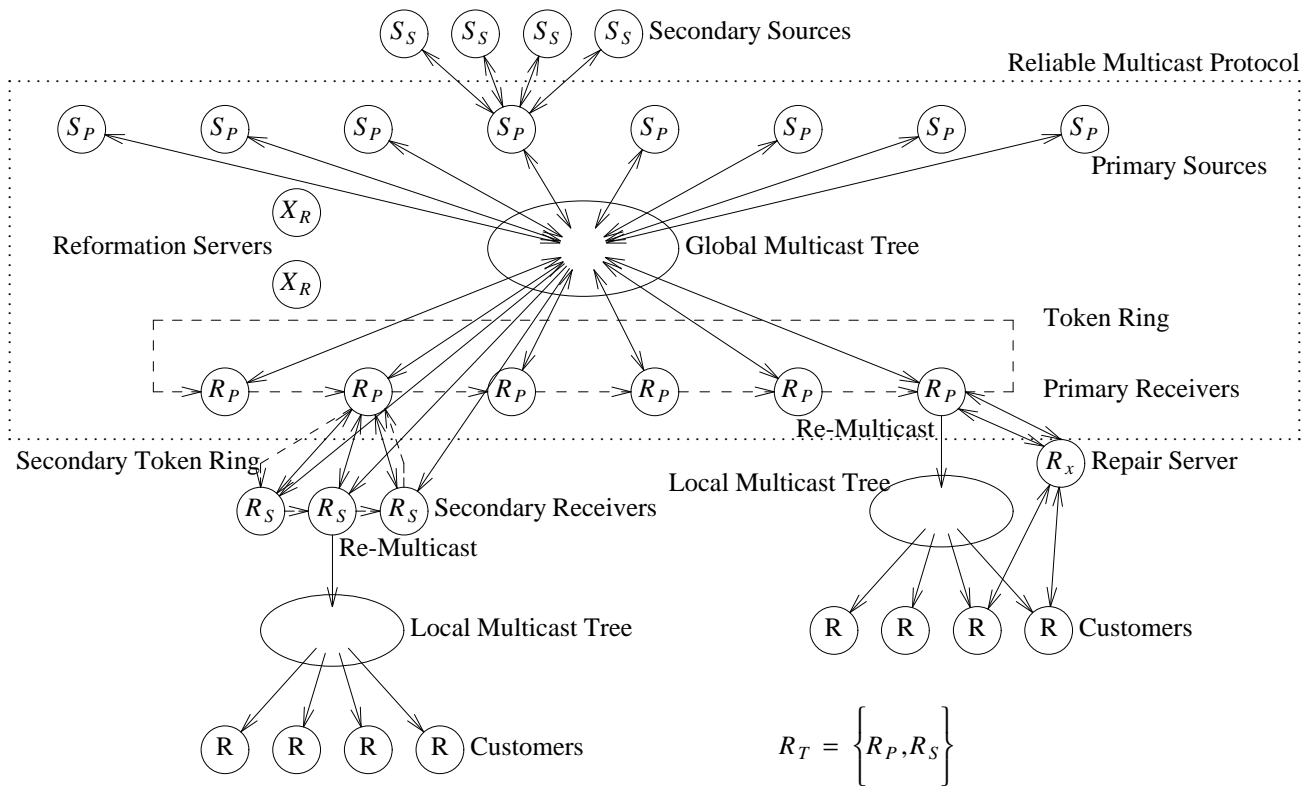

**Figure 2.** The general multicast architecture for the stock market applications.

### 4.1 Receivers

The primary receivers, $R_P$, take part in the RMP protocol. These components are trusted to share the information and not to favor specific customers. They can be owned and operated by the network provider or they can be licensed and regulated by a government organization. The set of receivers in the RMP protocol is stable and does not change as individual traders join or leave an application. Therefore, it is only necessary to reform the receiver group when a network based receiver fails and not when customers, $R$, join or leave the application.

The $R_P$ place the source messages in the same sequence, then multicast the sequence to the $R$. There is no limit on the number $R$ that can receive the multicast signal from an $R_P$. However, in order to meet our fairness conditions, the region of an $R_P$'s remulticast must be limited by the loss characteristics of the network and the delay from that $R_P$ to its $R$'s. In addition, multicast regions should overlap so that if an $R_P$ fails, or if the loss in an area of the network increases, an $R$ can obtain the information from an alternate source until the network is reformed.

Increasing $m$, decreases the size of the remulticast regions and can improve the quality of the data delivered to $R$. However, as $m$ increases, the time to detect failed $R_P$ and the time to guarantee that all of the receivers have an acknowledged message increases. If $m$ becomes very large the protocol becomes susceptible to NACK implosion because of the the correlated losses on the global multicast tree. We use a layered architecture, with secondary receivers, $R_S$, to improve the quality of the data delivered to $R$ without as large an increase in $m$.

The secondary receivers, $R_S$, are also trusted network servers. They receive the same source messages and acknowledgements as the $R_P$, and remulticast the same sequence of messages. However, the $R_S$ do not acknowledge source messages and do not take part in the primary token loop. When an $R_S$ detects a missing acknowledgement or source message it requests the message from a specific $R_P$ that is assigned to support it, rather than from the token site. Therefore, the $R_S$ do not increase the time required to detect a failed $R_P$ or add to the NACK implosion on the global multicast tree.

The $R_S$ in a region pass a secondary token between themselves to detect failures and guarantee that all of the source messages are received by all of the $R_{\bar{S}}$. The secondary tokens are numbered to correspond with the primary tokens and an $R_S$ does not pass the token until it has the acknowledgements and source messages up to that token number. A site that receives a secondary token can make the same inferences about the sites in the secondary group as a site that received the primary token could make about sites in the primary group. Each secondary token ring includes an $R_P$ on the primary token ring. That $R_P$ can infer state of the secondary group and transfer that information to the primary group.

As we increase the number of receivers in a secondary group we encounter the same problems that we encountered as we increased $m$. It takes longer to pass the token around the secondary group. Two layers of receivers can only reduce the number of token passes required to detect failures or guarantee delivery, $L_2$, by the square root of the number of passes in a single layer architecture, $L_1$. If we have a group of 100 receivers $L_1 = 100$. If we organize the 100 receivers into 10 secondary groups of 10 receivers, each secondary group requires 10 token passes to circulate the token and the primary group also has 10 members so that we require 10 token passes to detect a failure, $L_2 = 10$. If we have 10,000 receivers, $L_1 = 10,000$ and $L_2 = 100$.

We can generalize the layered receiver architecture. With $i$ layers, $L_i = \sqrt[i]{L_1}$. The land surface of the earth is about $57 x 10^6$ square miles. If we place 10,000 receivers uniformly over the surface, the maximum distance to a receiver is less than 50 miles, which is most likely adequate to provide our delay and loss guarantees. Since our stock market systems won't have to provide uniform access to the entire land surface of the earth, two layers of receivers should be more than sufficient to limit our token rings to a few 10's of receivers.

Within a region we can provide two grades of service, best effort and guaranteed delivery. Best effort delivers all of the messages that are not lost in the regional area to the traders. Guaranteed delivery recovers all of the messages. We guarantee delivery by colocating retransmit servers, $R_x$, with the $R_P$ or $R_{\bar{S}}$. When an $R$ detects a missing sequence number, it can request the message from $R_x$.

### 4.2 Sources

The sources, $S_P$, are trusted to enter valid data messages into the reliable multicast group. There are authentication[10,11,12] and certificate granting systems that can extend trust to a large number of sources, however, we can keep tighter security with a smaller number of participants. In addition, RMP requires that each receiver maintain state information about each source and some of our cryptographic techniques require receivers to share a secret with each source. As the number of sources becomes large, it becomes difficult to maintain source specific information.

While the constraint on the number of sources isn't as severe as the constraint on the number of receivers, there should not be millions or ten's of millions of sources in an RMP group. In order to keep the number of sources that participate in RMP to a few thousand, the sources can also be layered: The $S_P$ participate in RMP; The $S_P$ trust a set of secondary brokers, $S_S$; And, each $S_S$ trusts a set of customers.

## 4.3  Reformation

In a positive acknowledgement protocol the source assumes that the receiver has failed if it does not receive an acknowledgement after a specified number of transmission attempts. Token group failures are detected when an $S_P$ fails to receive an acknowledgement for a message, a token site fails to pass a token, or a receiver cannot recover a missing message or acknowledgement. When a site detects a failure it notifies a reformation server, $X_R$. There are redundant reformation servers in case one fails.

The $X_R$ is responsible for forming new primary and secondary token rings in the groups that have failures. It also assigns addresses for the remulticasts from the $R_T$ (the set of $R_P$ and $R_S$) and adjusts the scope of their multicasts so that every customer can receive from at least two multicast sources but two multicast sources that use the same address do not overlap.

In the original RBP protocol the reformation process was distributed, instead of centralized, and every operable receiver was equally likely to become the reformation server. The centralized protocol is simpler because it does not have an election phase to decide which server becomes $X_R$. In the stock market application the reformation is more complicated because it may be limited to a subgroup of receivers on particular secondary rings. In addition, the $X_R$ needs to have more information because the reformation operation includes adjusting multicast regions and possibly changing the remulticast addresses. The additional information on receiver group structures, multicast ranges, etc. is better maintained in a small number of $X_R$ than in possibly 100,000 $R_T$'s.

To determine the operable receivers and reform a token group, the $X_R$ multicasts several invitations to join a new group of receivers. The operable receivers use a positive acknowledgment protocol to send a message to $X_R$. If the receiver does not receive the acknowledgement it decides that the $X_R$ failed and tries another $X_R$. The message to $X_R$ includes the number of the receiver's last acknowledged message, $t_r$. The reformation server creates a new token loop, decides on the starting message number $t$, and appoints a receiver with $t_r \geq t$ as the token site.

The starting message number, $t$, for the new repair group depends on the application. In the unified ticker the token can start at $t = t_{r,\max} + 1$. In a distributed trading floor we may want to be more conservative, as we were in the distributed database on RBP,[2] to make certain that trades aren't lost when sites fail or recover.

## 4.4  Striping

The amount of data in the composite stock ticker will almost always exceed the bit rate that can be transmitted to an individual customer, $R$. The amount of data in the composite ticker may also exceed the amount of data that can be processed by an $R_T$ in the repair group. Both of these problems are addressed by "striping" stocks into common groups. However, the stripes that are used to solve the two problems have different widths.

Instead of having one set of $R_T$ that process all of the messages, stocks are organized in stripes of related stocks. Each stripe uses a different multicast address for the RMP protocol at the core of the architecture and uses a different set $R_T$ to process the messages. A stripe is limited to a group of stocks that have few enough messages to be processed by all of the $R_T$. Since the $R_T$ are provided by the network, rather than the customers, we can assume that all of the processors are similar - no weak links - and that they are among the more powerful processors that are available. There should be a relatively small number of wide stripes in the core of the architecture.

An $R_T$ organizes the data on the stripe that it receives into narrower stripes of more closely related stocks and transmits each stripe on a different remulticast addresses. The amount of data on one of the final stripes is limited by the minimum reception rate of the $R$. Any of the $R$ in the remulticast region can receive the data on one stripe and an $R$ with higher rate lines can receive the data on multiple stripes. For instance, if the least capable $R$ have 56 Kbps modems, $R_T$ organizes the data into 56 Kbps stripes. A customer with a 56 Kbps modem may select any one remulticast address and receive information on a small group of stocks, while a broker with a 45 Mbps connection may simultaneously view the information on the stocks in 800 different stripes. There should be a relatively large number of narrow stripes at the edge of the architecture.

## 5. Protocol Modifications

The fairness issues and the quantity of data in the stock market applications require modifications of the basic RMP protocol. The major changes are

1.  delayed delivery,

2.  time driven, rather than event driven, token passing, and

3.  NACK reduction.

In the modified protocol, the $R_T$ wait $\Delta_A$ between the time a message is acknowledged and the time that it is remulticast. The $R_T$ have synchronized clocks. The clocks can be synchronized by a network protocol[13,14], or by receiving a signal from a satellite and adjusting for differences in delay. When the token site acknowledges a source message, it timestamps the acknowledgement. The $R_T$ remulticast a message $\Delta_A$ after the message's time stamp. The delay $\Delta_A$ compensates for the difference in reception times at the different $R_T$ caused by differences in propagation delay from the source and the time to detect and recover missing messages. A system is fair when all of the $R_T$ simultaneously retransmit the message to their local customers.

In the modified protocol, the token is transferred periodically, every $\tau_t$ seconds, and acknowledges all of the unacknowledged source messages, including any messages that were missed by the previous token sites. The $t^{th}$ token passing message can acknowledge a sequence of $k$ source messages, where $k$ is variable. The messages are assigned sequence numbers $s$ to $s + k - 1$.

In section 5.1 we show that the maximum token transfer rate may not be as fast as the message arrival rate in some of our applications. When this occurs in the original RMP protocol the queue of messages that are waiting to be acknowledged grows and the latency before messages are acknowledged can become large. Acknowledging multiple messages prevents this from happening.

We can acknowledge multiple source messages without passing the token periodically. In an event driven protocol we can acknowledge all of the waiting messages, or wait for the next message when there are no messages waiting. An event driven protocol transmits fewer tokens when the message arrival rate is low and acknowledges source messages sooner when the message arrival rate is high.

However, periodic token transfers have compelling advantages for the stock market applications. To start with, periodic token transfers provide more deterministic times to recover missing messages, detect failed receivers and to guarantee that all of the sites have received a message. The times are more deterministic because we detect missing acknowledgments when their scheduled transmission time passes, instead of waiting until we receive the next acknowledgement. The next acknowledgement is initiated by the arrival of the next source message, which is a random variable that may be long relative to the token transfer time.

In section 5.1 we show that a periodic token transfer system is fair, all or the $R_T$ remulticast them messages at the same time, when $\Delta_A > \tau_t$, just one token transfer time. In the original RMP protocol there is a relationship between the number of token passes, up to a complete cycle, and the probability that a receiver has a message. When we delay delivery, we can guarantee a higher probability of simultaneous remulticast, but cannot guarantee fairness.

A secondary advantage of periodic token transfers is that the other $R_T$ detect a failed token site when a token isn't transferred on time. We do not have to wait for a source to complain that it is unable to elicit an acknowledgement. Removing failure detection and reporting responsibility from the sources makes it possible to operate with less trusted sources. In section 7 we show that this can be an important characteristic in an environment where all trading floors are not equally trusted.

NACK implosion is considered an important problem in reliable multicast systems. RMP has mechanisms to reduce NACK implosion that are not available to other reliable multicast protocols. Our architecture reduces NACK implosion to the point where it is not a problem, and the NACK reduction mechanisms should not be used in the current stock market applications. In section 5.3 we describe the reduction mechanisms in case NACK implosion becomes a problem in a future system.

## 5.1 Delay - Fairness and Latency

The original RMP protocol was event driven. $\Delta_A$ seconds after an acknowledgement is transmitted there is a probability that a receiver that missed the acknowledgement is still unaware of the fact that the acknowledgement is missing. The probability that the receiver is unaware of a missed acknowledgement $\Delta_A$ seconds after it occurs is a function of the number of additional acknowledgements that have been transmitted. A receiver must miss all of the subsequent acknowledgements to remain unaware of the original loss. The number of additional acknowledgements that are transmitted in an interval of $\Delta_A$ seconds is a variable, and may be zero. The longer we wait, the more likely it is that a receiver detects and recovers a missing message. Therefore, there is a tradeoff between delay and fairness.

In the modified, time driven protocol we can make a much stronger claim. We can state that, when we make $\Delta_A$ sufficiently large then all operable receivers will recover the message before $\Delta_A$ or the system will be in reformation. This statement isn't as strong as the guarantees that are provided by circulating the token around the entire loop, but we will show that $\Delta_A$ is much smaller than the time to circulate the token.

The token passing message is scheduled at a particular time. Make $\Delta_A$ long enough for a receiver to either detect the missing token passing message, based upon its late arrival, recover that message, then recover any missing source messages that were acknowledged, or start a reformation process. When $\Delta_A$ is at least this long, every $R_T$ either has the acknowledged message and remulticasts it simultaneously or the system is being reformed.

At $\delta_N$ after a scheduled token passing message, receivers infer that they have missed the token passing message and request a retransmission. The delay in packet networks is a random variable and can be very long. $\delta_N$ is only long enough to guarantee that a reasonable fraction of the messages, not all of the messages, are received. The penalty for not making $\delta_N$ long enough to receive all of the messages are extra requests and retransmissions.

Both the token passing messages and source messages are recovered with a positive acknowledgement protocol. We resend the request at regular intervals, $\tau_r$, and declare the system inoperable if after $k_r$ attempts the request is not answered. Therefore, after $\delta_N + 2 k_r \tau_r$ a receiver either has the source message that was acknowledged, or has declared the system inoperable and asked for a reformation. The interval between requests is at least $2 \delta_N + X + P$, where $X$ is the time to transmit the request and response and $P$ is the time to process the request and response.

There is an underlying assumption that token sites transmit the token on time. In order for the token site to always have the token before it is scheduled to transfer it, the time between token transfers, $\tau_t$, must be long enough for the token site to determine that the token transfer message was lost, recover that message, then recover any source messages that are missing. Therefore, $\tau_t \geq \delta_N + 2 k_r \tau_r$ ($= \Delta_A$). The system is completely fair as long as the $R_T$ wait one token passing time after the message is timestamped before remulticasting it.

The token passing time can be calculated as $\tau_t = (1+2k_r)\delta_N + k_r(X+P)$. $\delta_N$ is the sum of the propagation delay between a source and receiver plus the router delays for our "reasonable fraction of the messages." The circumference of the earth is about 25,000 miles. Therefore, $\delta_N$ should be on the order of 75 milliseconds to allow the signal to propagate half way around the world. The backbone of the stock market network, where RMP is implemented, has dedicated bandwidth and is not subject to the large delays that occur in the public Internet. In addition, control messages should have priority over new source messages. The "reasonable" router delay on a path can probably be kept to a few hundred milliseconds. The processors in the network are dedicated to this application. Therefore, $P$ can be kept to a few 10's of milliseconds. The source and control messages in the system are short, relative to file transfers, and the backbone is a high rate, multi-megabit per second transmission facility. Therefore, $X$ is negligible. In typical protocols, $k_r = 3$. With these parameters, the token transfer time is about 3 seconds. In stock market systems we expect many more than one new source message every 3 seconds, particularly during peak periods. This shows the need to acknowledge multiple source messages with the same token passing message.

## 5.2 NACK Reduction

The standard mechanism for reducing NACK implosion in reliable multicast systems is to limit the subset of receivers that request a missing message, but to multicast the missing message to all of the receivers. In subsequent intervals of time different subsets of receivers can request the missing message until all of the receivers have had the opportunity. If a receiver requests a message and a receiver that is scheduled to request the message in a later interval receives the multicast, the later receiver does not request the retransmission, and the number of NACK's is reduced. This strategy is particularly appropriate in the Internet where the multicast is transmitted on a tree and many receivers miss the same message.

In the RMP protocol the receiver that accepts the token must have all of the source message. Therefore, when we define subsets of receivers that request missing messages, we must guarantee that a receiver has an opportunity to recover a missing message before it becomes the token site.

The simplest and most economical method of defining subsets in RMP is to have one receiver in each subset, the next token site. A receiver can only request missing messages before it becomes the token site, but, before becoming the token site, can recover any message that was also missed by a preceding token site. The disadvantage with this approach is that a site may have to wait an entire token rotation before it can recover a missing message.

In order to reduce the time until a site can recover a missing message, we give several sites the opportunity to request the missing message. We space those sites equidistant around the token ring to minimize the maximum time until a receiver can request a missing message. If token $t$ is sent by receiver $r$, define sets of receivers $S_{i,t} = \left\{ (r+i+1+j*k_p) \bmod m \text{ for } 0 \le j \le (m-i-1)/k_p \right\}$, for $i = 0, 1, ..., k_p - 1$, where there are $m$ receivers numbered 0 to $m-1$ in the token group. After a receiver in $S_{i,t}$ receives token $t+i$ it can request the message acknowledged by $t$, if it is still missing. With this assignment, every receiver can request a missing message within $k_p$ token passes.

If $m/k_p$ is an integer, each receiver requests any missing messages in the interval it is scheduled to accept the token, and every $k_p^{th}$ interval after that. In the other $(k_p - 1)^{th}$ intervals, the receiver listens in case one of its missing messages is recovered by one of the other receivers.

Since the receivers in the later sets do not request a missing message if a receiver in an earlier set requests the same message and the retransmission is received, the average number of requests for retransmission is clearly reduced. If, on the average, there are more sites that miss the message than there are sets, we can further reduce the average number of requests by putting fewer receivers in the sets that make the initial requests than in the sets that make later requests. We should "tune" the number of receivers in each set so that, on the average, the probability of a request is the same in each subset. We can also reduce the number of requests by placing receivers in different sets if they are likely, because of their positions on the original multicast tree, to miss the same messages.

A problem with limiting the number of receivers that transmit a NACK is that it increases the delay until a missing message may be acquired. In the stock market application we have to wait for $k_p$ token passes, instead of one, for complete fairness. Since NACK reduction increases $\Delta_A$ in the stock market applications, we do not recommend it. In a different application we may replace fairness with a penalty for recovering the message later. In this type of an application, NACK reduction may be beneficial.

## 6. Security

The security issues are among the biggest differences between the the Internet stock market application and the local network database. In section 3 we identified the following security concerns:

1. constraining transmission access to authorized sources,

2. preventing early reception of the data stream,

3. limiting reception to authorized receivers,

4. spoofing or adding to the repaired sequence, and

5. denial of service.

The global multicast group, $G_M$, includes $S_P$, $R_P$, $R_S$ and $R_x$. Our first security concern is that a source outside our set will transmit messages that are placed in the sequence of messages that are remulticast. The messages in $G_M$ are delayed before they are remulticast to $R$ so that all of the receivers get the messages simultaneously. The members of the $R_T$ are trusted not to divulge the messages early. Our second concern is that an unauthorized receiver will eavesdrop on $G_M$.

We address the first two concerns with a combination of cryptographic and networking techniques. In our architecture there are only a few hundred $S_P$. We can operate $G_M$ on a public network if each $S_P$ shares a secret key with the group of $R_T$ and encrypt its messages. If each $S_P$ has a different key, the encryption also identifies the source, and hides messages from other sources. We can also operate $G_M$ on a private network, and use firewalls to prevent unauthorized transmitters and receivers.

The networking approach has the advantage that the bandwidth of the backbone network is not shared with the public network, and can be guaranteed. In addition, a network with restricted access reduces the threat of denial of service attacks since those with legitimate access to the resources have obtained some type of approval. A cryptographic approach, with a separate secret for each source, has the advantage that the $S_P$ are not trusted to the same degree as when they transmit in the clear on a private network. We do not have to trust the sources to refrain from using or redistributing the messages from other sources. $G_M$ can use both mechanisms: A private network for the backbone multicast and a shared secret that is different for each source.

Our third concern is to restrict access to the data that is remulticast by the $R_T$. There are electronic stock markets, like NASDAQ, that require the $R$ to be part of a private network. Our objective is to make our system accessible to the general population, less expensively, by using the public Internet to connect the $R$. This does not prevent some participants from having dedicated connections.

An $R_T$, ISP, or stock market can give the remulticast sequence away for free, to attract customers, or the sequence can be sold by the month, like a subscription for a newspaper. In order to sell the multicast sequence on the public Internet, the $R_T$ decrypt the messages from the $S_T$, then re-encrypt the entire sequence with a new key. The key is sold to each of the receivers, $R$, and is changed when the subscriptions expire.

The problem of selling and distributing the remulticast key is similar to the problems encountered in other Internet distribution systems. The information has limited value, in that the key can be bought. However, we must discourage someone who buys the key from giving it to others. This problem has been addressed in an earlier electronic publishing experiment that also used Internet multicast[15,16]. In that experiment the key was included in a program that was sent to each subscriber when they paid for the service with their credit card. The key is masked by the credit card number, so that a recipient who gives away a copy of the program must also give away his credit card number in order for the program to be useful. While this does not prevent a person from giving away the program, it should discourage him.

The final two security concerns occur on the remulticast groups on the public Internet. A malicious user may pretend to be the remulticast receiver and insert false messages into the sequence, or may flood the multicast group to prevent others from receiving the repaired sequence.

The conventional cryptographic approach for dealing with pretenders is digital signatures. Since there are a large number of untrusted receivers, the digital signature should use public key cryptography[17,18]. In a public key system only the remulticast source can sign the message, but any receiver can verify that the message is legally signed. Public key systems require more computation than we would like to perform in this application.

The messages in our system are numbered. We assume that an attacker cannot remove multicast messages. We can therefore detect that messages have been added to the sequence. We cannot tell which of the duplicate messages are real, but we can decide that some messages are forgeries and not act on any of the information.

Flooding the multicast group to deny service to the receivers cannot be prevented by cryptographic techniques. Since there is only one source in our multicast group, we can eliminate illegal transmitters by configuring routers to only multicast the signal from a particular source on a particular port. In addition, assuming that flooding is illegal or at least discouraged by ISP's, it can eventually be stopped.

## 7. Applications

The three applications of the RMP protocol are

1. a unified ticker of the transactions on the physical and electronic trading floors,

2. a merged stream of buy and sell orders, and

3. a distributed trading floor.

The first application has a relatively small number of sources and a very large number of receivers. The second application has a very large number of sources and a relatively small number of receivers. The third application has the same number of sources and receivers, both of which may be large.

In the first two applications we are primarily interested in fairness. RMP is used to create a level playing field for investors independent of their location. In the third application we are also interested in the protocol's ability to provide the same sequence of messages to every receiver, so that receivers can independently determine which transactions have occurred. we are also interested in the protocol's ability to only commit transactions after a specified number of receivers have witnessed them, so that the transactions can survive system failures.

### 7.1 Unified Ticker:

In the unified ticker every investor receives a list of the trades on every trading floor. The objective is to create a level playing field where all of the investors have the same information on trades. The investors receive the list in the same order, at the same time, no matter where in the world they are located.

The sources in the RMP architecture, $S_P$, are the trading floors. The trading floors operate independently under their own rules and customs. Some may be physical places, others may be Internet servers, and still others may be the distributed trading floors described in the third application. Investors know the rules and risks associated with the different trading floors and contact them directly or through a broker.

There are a few hundred trading floors and they submit their list of trades directly to the the RMP multicast group. The core multicast network is a virtual private network and the trading floors are inside a firewall. The trading floors are trusted to a certain extent. They are trusted to place only an honest and timely accounting of their trades into the multicast sequence, and not to divulge their own trades before they are reported on the unified ticker. We assume that this degree of cooperation can be enforced by the regulatory agencies or by the fear of being excluded from the unified ticker.

Each trading floor shares a secret key with the group of receivers, $R_T$, and encrypts the messages that it places in the multicast sequence. The encryption serves two functions: Firstly, it acts as a signature of the trading floor that has entered the data. Secondly, it reduces the degree of trust in trading floors, since trading floors cannot acquire and use the information before it is available on the unified ticker. The second function is particularly important because the trading floors are not all equals and different floors may be under different international regulatory agencies, with different rules and penalties.

The multicast receivers, $R_T$, are owned by the network operator. The primary receivers, $R_P$, acknowledge and time stamp the source messages. At $\Delta_A$ after the timestamp, on the order of 3 seconds in an international system, either all of the $R_T$ have the acknowledged message, or the network has malfunctioned and gone into the reformation process. All of the $R_T$ decrypt a source message at the same time and remulticast that message in their regional areas.

The regional multicasts may be outside the firewalls of the virtual private network. If the unified ticker is being sold, rather than provided as a free service to the receivers in a region or the customers of an ISP, it is re-encrypted with a secret key that is different from the key that is used by any of the sources. This key is shared with all of the receivers in a region, as described in the previous section. The multicast in the regional area is susceptible to spoofing or denial of service attacks. The options for dealing with these threats are also described in section 6.

The messages in the regional multicast may be lost. A receiver, $R$, can detect a lost message by a missing sequence number. We can colocate a retransmit server, $R_x$, at an $R_T$. When $R$ detects a missing sequence number, it requests the message from $R_x$. The data transmitted in a unified ticker is temporal and many customers may only be interested in the most recent stock prices. Once the next value of a stock is received, there is no reason for these customers to retrieve the previous price. There may be other customers who wish to accurately plot the stock price to predict trends. These customers may retrieve the missing transactions.

There is an expense associated with maintaining $R_x$. Network provider can recover the expense by selling two levels of service, one with and one without retransmissions. It is likely that the same messages will be lost by many receivers in a region. Therefore, retransmissions should be multicast. The retransmissions can be encrypted with a different key than the original multicast. The retransmit key can be sold separately by the same techniques as the multicast key. Only receivers that pay for the higher level of service can decrypt the retransmitted messages. An ISP may also give away the base service and charge for the more reliable service.

The amount of data in a unified stock ticker will be significantly greater than the amount of data that a typical user can receive. This problem is addressed by striping. If the base rate of the receivers is 56 Kbps, similar stocks are joined in stripes that are unlikely to require a data rate exceeding 56 Kbps. Each stripe is transmitted on a different multicast address. A trader with a 56 Kbps modem can only select one stripe at a time while a trader with a 1.5 Mbps line may simultaneously follow the stocks on 25 stripes.

Striping, with much wider stripes, may also be necessary in the backbone network. In this case, the entire multicast infrastructure, $R_P$'s, $R_S$'s, and $R_x$'s, is duplicated for each stripe. The $S_P$'s transmit transactions involving different stocks on the appropriate stripe. We assume that network bandwidth will not constrain the stock transactions that can be reported, as in the end system, and that the stripes are needed because the message rate exceeds the throughput of the $R_T$.

## 7.2 Unified Orders

The unified order system is a sequence of offers to buy or sell stocks at a given price. The offers can be directed to a specific exchange or can be open to all participating exchanges. Our objective is to give all of the traders a fair opportunity to place their bids in the sequence of offers.

If the buy and sell offers are directed to a single exchange, the order of the sequence may be binding on the trades that occur. If the offer is open to all exchanges, the offer may just be an invitation for a broker to close a deal. An open offer to all exchanges may also be an order that is too large to be handled by a single exchange.

This system is the inverse of the unified ticker. There are many sources and a small number of receivers. In the degenerate case there is one receiver, a single trading floor. It may seem wasteful to circulate the token among the $R_P$ that are distributed around the world, just to give the sequence to a single trading floor in a single location. However, the circulating token is needed for fairness.

If all of the sources transmit their offers directly to the trading floor, the sources that are in the same city as the exchange have an advantage over sources on the other side of the world. Firstly, the propagation and network delays may be seconds shorter, and secondly, the average number of trys to have the message correctly received may be several times smaller. If two traders in different parts of the world receive the unified ticker and decide simultaneously to sell their stocks at the same price, the trader in the same city as the trading floor will almost always have its order registered first.

With a rotating token, the portal that allows messages to enter the system spends equal amounts of time at different locations on the globe. The trader that gets into the system first depends on where the portal is located when the two traders decide to enter their offers, and not where the trading floor is located.

The sources in this application include brokers and individual traders as well as trading floors. These sources cannot be trusted to the same extent as the trading floors in the unified ticker. The sources may make offers without the proper resources, or may transmit a large number of messages to disrupt the system. In addition, there may be too many sources for the $R_T$ to have a different shared secret with each.

Both of these problems are solved by not giving the sources direct access to the RMP group. The sources, $S_P$, are either owned by the network or are completely trusted. These are the only sources inside the network firewall. The sources, $S$, must present credentials to the $S_P$ that they own the stock that they would like to sell or that they have the funds that they would like to spend. Alternatively, the $S$ may have accounts with the $S_P$, in which case they must prove their identity by an agreed upon password system. If there are too many $S$ for the network based $S_P$ to track, there can be secondary sources, $S_S$, that trust the $S$ and are trusted by the $S_P$.

## 7.3 Distributed Trading Floor

The third application uses RMP to construct a distributed, international trading floor. The participants may be individual traders, brokers, or the other trading floors. This trading floor may be one of the sources in the unified ticker. All of the participants may enter buy, sell or stop orders and must see all of the orders. Depending on the sequence of the orders, each participant knows which trades have occurred.

This application has many of the problems of the previous two applications. There are large numbers of sources and receivers, none of which is trusted. Both the $S_P$ and $R_T$ must be network based. They are distributed around the world to provide fair entry and distribution of the data. The $S_P$ verify the credentials of the $S$ and enter the bids. Based upon the RMP sequence and the rules of the particular trading floor, an arbiter declares trades to be made and reports the trade on an appropriate ticker. By making the token site the arbiter we can guarantee that the arbiter has the most complete acknowledged sequence of buy and sell orders.

There are a number of different rules that we can use to make trades. Some rules are semantic. If one participant offers to buy a stock at price $A$ and another offer to sell the stock at price $B < A$, should the trade be made at $A$, $B$, or somewhere in between. Other rules are a matter of style. Some floors may post buys and sells, others may be run a single round, high bid auction, where the highest bidder gets to buy the stock at the price offered by the second highest bidder. Other floors can be modeled after the Amsterdam flower auction.

RMP offers very strong guarantees that can be used to make trades reliably even when there are system failures. For instance, assume that the token site is the arbiter. An $R_P$ that has the token can report a tentative trade as soon as it sees a match. The next time that this $R_P$ receives the token it can report that the trade is confirmed because every $R_P$ has received the report. The trade information cannot be lost as long as one of the $R_P$ survives.

## 8. Conclusions

We have described an architecture and protocols for an Internet-based global stock exchange. RMP ensures fair and timely entry and delivery of data, and scales globally.